\documentclass[12pt]{iopart}
\usepackage[dvipdfm]{graphicx}
\usepackage{iopams}
\usepackage{float}
\newcommand{\pa}{\partial}
\newcommand{\al}{\alpha}
\newcommand{\del}{\delta}

\begin{document}
\title[Image formation in gravitational lensing]{Wave optics and image
  formation in gravitational lensing}
\author{Yasusada Nambu}
\address{Department of Physics, Graduate School of Science, Nagoya 
University, Chikusa, Nagoya 464-8602, Japan}
\ead{nambu@gravity.phys.nagoya-u.ac.jp}
\date{July 30, 2012, ver. 0.97} 
\begin{abstract}
  We discuss image formation in gravitational lensing systems using
  wave optics. Applying the Fresnel-Kirchhoff diffraction formula to
  waves scattered by a gravitational potential of a lens object, we
  demonstrate how images of source objects are obtained directly
  from wave functions without using a lens equation for gravitational lensing.
\end{abstract}
\pacs{04.20.-q, 42.25.Fx}
\vspace{2pc}
\noindent{\it wave optics; image formation; gravitational lens}

\maketitle

\section{Introduction}

Gravitational lensing is one of the prediction of Einstein's general
theory of relativity and many samples of images caused by
gravitational lensing have been obtained
observationally~\cite{SchneiderP:SV:1992}.  Light rays obey null
geodesics in curved spacetime and they are deflected by gravitational
potential of lens objects.  In weak gravitational field with thin lens
approximation, a path of a light ray obeys so called lens equation for
gravitational lensing and many analysis concerning the gravitational
lensing effect are carried out based on this equation.  Especially, we
can obtain images of source objects by solving the lens equation using
a ray tracing method.  As a path of light ray is derived as the high
frequency limit of electromagnetic wave, wave effects of gravitational
lensing become important when the wavelength is not so much smaller
than the size of lens objects and in such a situation, we must take
into account of wave effects.  For example, when we consider
gravitational wave is scattered by gravitational lens objects, the
wave effect gives significant impact on the amplification factor of
intensity for
waves~\cite{NakamuraTT:PTPS133:1999,BaraldoC:PRD59:1999,MatsunagaN:JCAP01:2006}.
Another example is direct detection of black holes via imaging their
shadows~\cite{FalckeH:AJ528:2000,MiyoshiM:PTPS155:2004}. The apparent
angular size of black hole shadows are so small that their
detectability depends on angular resolution of telescopes which is
determined by diffraction limit of image formation system. Thus it is
important to investigate wave effects on images for successful
detection of black hole shadows.

Although interference and diffraction of waves by gravitational lensing has been
discussed in connection with amplification of waves, a little was
discussed about how images by gravitational lensing are obtained based
on wave optics. For electromagnetic wave, E.~Herlt and
H. Stephani~\cite{HerltE:IJTP15:1976} discussed the position of images
by a spherical gravitational lens evaluating the Poynting flux of
scattered wave at an observer.  They claimed that there is a
disagreement between wave optics and geometrical optics concerning the
position of double images of a point source. But they have not presented
complete understanding of image formation.  In this paper, we
consider image formation in gravitational lensing using wave optics
and aim to understand how images by gravitational lensing are obtained
in terms of waves.
For this purpose, we adopt the diffraction theory of image formation
in wave optics~\cite{SharmaKK:AP:2006}, which explains image formation
in optical systems in terms of diffraction of waves. 
\section{Wave optics in gravitational lensing}
We review the basic formalism of gravitational lensing based on wave
optics~\cite{SchneiderP:SV:1992}. In this paper, we does not consider
polarization of waves and treat scalar waves as a model for
electromagnetic waves.
\begin{figure}[H]
  \centering
  \includegraphics[width=0.8\linewidth,clip]{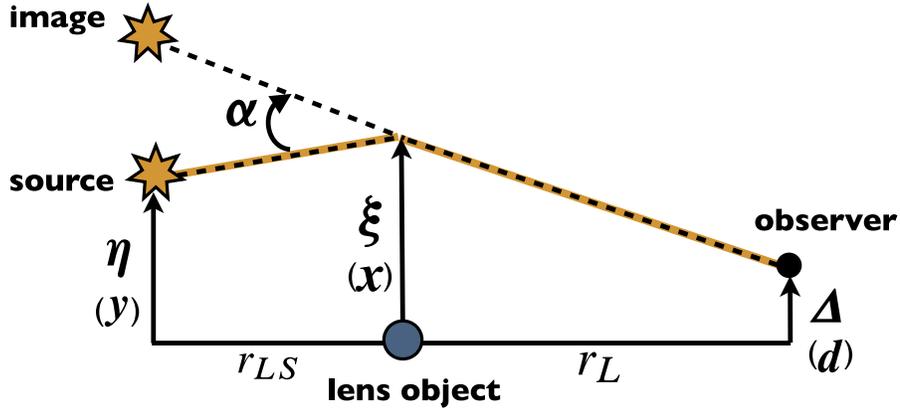}
  \caption{The gravitational lens geometry of the source, the lens and
    the observer. $\al\ll1$ is the deflection angle. $r_L$ is the
    distance from the observer to the lens object and $r_S$ is the
    distance from the observer to the source.  $r_{LS}=r_S-r_L$.}
  \label{fig:glens1}
\end{figure}
Let us consider waves propagating under the influence of the
gravitational potential of a lens object. The background metric is
assumed to be
\begin{equation}
  ds^2=g_{\mu\nu}dx^\mu dx^\nu=-(1+2U(\boldsymbol{r}))dt^2
  +(1-2U(\boldsymbol{r}))d\boldsymbol{r}^2,
\end{equation}
where $U(\boldsymbol{r})$ is the gravitational potential of the lens object
with the condition $|U|\ll1$.  The scalar wave propagation in this
curved spacetime is described by the following wave equation:
\begin{equation}
  \pa_\mu\left(\sqrt{-g}g^{\mu\nu}\pa_\nu\Phi\right)=0,
\end{equation}
and for a monochromatic wave with the angular frequency $\omega$,
\begin{equation}
  (\nabla^2+\omega^2)\Phi=4\omega^2U(\boldsymbol{r})\Phi,
\end{equation}
where $\nabla^2$ is the flat space Laplacian. 

We show the configuration of the gravitational lensing system
considering here (Figure~\ref{fig:glens1}). The wave is emitted by a
point source, scattered by the gravitational potential of the lens
object and reaches the observer. We assume the wave scattering occurs
in a small spatial region around the lens object and outside of this
region, the wave propagates in a flat space. With the assumptions of
the eikonal and the thin lens approximation, the Fresnel-Kirchhoff
diffraction formula provides the following amplitude of the wave at
the observer~\cite{SchneiderP:SV:1992,BaraldoC:PRD59:1999}
\begin{equation}
  \Phi(\boldsymbol{\eta},\boldsymbol{\Delta})=\frac{\omega a_0}{2\pi ir_{LS}r_L}\int
  d^2\xi\exp\left[i\omega S(\boldsymbol{\eta},\boldsymbol{\xi},\boldsymbol{\Delta})\right]
\end{equation}
where $S(\boldsymbol{\eta},\boldsymbol{\xi},\boldsymbol{\Delta})$ is
the effective path length (eikonal) along a path from the source
position $\boldsymbol{\eta}$ to the observer position
$\boldsymbol{\Delta}$ via a point $\boldsymbol{\xi}$ on the lens plane
\begin{eqnarray}
  S(\boldsymbol{\eta},\boldsymbol{\xi},\boldsymbol{\Delta})
  &=\left[(\boldsymbol{\xi}-\boldsymbol{\eta})^2+r_{LS}^2\right]^{1/2}
  +\left[(\boldsymbol{\xi}-\boldsymbol{\Delta})^2+r_L^2\right]^{1/2}
  -\hat\psi(\boldsymbol{\xi})
  \nonumber \\
  &\approx
  \frac{(\boldsymbol{\eta}-\boldsymbol{\Delta})^2}{2r_S}+r_S+\frac{r_Lr_S}{2r_{LS}}\left(
    \frac{\boldsymbol{\xi}-\boldsymbol{\Delta}}{r_L}-\frac{\boldsymbol{\eta}
      -\boldsymbol{\Delta}}{r_S}\right)^2-\hat\psi
  (\boldsymbol{\xi})
\end{eqnarray}
and we assume that $|\boldsymbol{\eta}-\boldsymbol{\Delta}|\ll r_S$ and
$|\boldsymbol{\xi}-\boldsymbol{\Delta}|\ll r_L$. A constant $a_0$ represents the
intensity of a point source.  The two dimensional gravitational
potential is introduced by
\begin{equation}
  \hat\psi(\boldsymbol{\xi})=2\int_{-\infty}^{\infty}dz U(\boldsymbol{\xi},z).
\end{equation}
 Then the wave amplitude at the
observer can be written as~\cite{SchneiderP:SV:1992,NakamuraTT:PTPS133:1999}
\begin{equation}
  \label{eq:phi}
  \Phi(\boldsymbol{\eta},\boldsymbol{\Delta})
  =\Phi_0(\boldsymbol{\eta},\boldsymbol{\Delta})F(\boldsymbol{\eta},\boldsymbol{\Delta})
\end{equation}
where $\Phi_0$ is the wave amplitude at the observer in the absence of the
gravitational potential $U$:
\begin{equation}
  \label{eq:phi0}
  \Phi_0(\boldsymbol{\eta},\boldsymbol{\Delta})
=\frac{a_0}{r_S}\exp\left[i\omega S_0(\boldsymbol{\eta},\boldsymbol{\Delta})\right], 
\qquad S_0(\boldsymbol{\eta},\boldsymbol{\Delta})
=\frac{(\boldsymbol{\eta}-\boldsymbol{\Delta})^2}{2r_S}
+r_S.  
\end{equation}
$S_0(\boldsymbol{\eta},\boldsymbol{\Delta})$ is the path length along a straight path
from $\boldsymbol{\eta}$ to $\boldsymbol{\Delta}$.  The amplification factor $F$ is
given by the following form of a diffraction integral
\begin{eqnarray}
  \label{eq:FF}
  &F(\boldsymbol{\eta},\boldsymbol{\Delta})
  =\frac{r_S}{r_Lr_{LS}}\frac{\omega}{2\pi i}\int
  d^2\xi\exp\left[i\omega S_1(\boldsymbol{\eta},\boldsymbol{\xi},
    \boldsymbol{\Delta})\right],\\
  &\qquad S_1(\boldsymbol{\eta},\boldsymbol{\xi},\boldsymbol{\Delta})
=\frac{r_Lr_S}{2r_{LS}}\left(\frac{\boldsymbol{\xi}-\boldsymbol{\Delta}}{r_L}
        -\frac{\boldsymbol{\eta}-\boldsymbol{\Delta}}{r_S}\right)^2
      -\hat\psi(\boldsymbol{\xi}) 
\end{eqnarray}
where $S_1(\boldsymbol{\eta},\boldsymbol{\xi},\boldsymbol{\Delta})$ is
the Fermat potential along a path from the source position $\boldsymbol{\eta}$
to the observer position $\boldsymbol{\Delta}$ via a point $\boldsymbol{\xi}$ on the
lens plane.  The first term in $S_1$ is the difference of the
geometric time delay between a straight path from the source to the
observer and a deflected path. The second term is the time delay due
to the gravitational potential of the lens object.  Now we introduce
the following dimensionless variables:
\begin{equation}
 \fl\quad
 \boldsymbol{x}=\frac{\boldsymbol{\xi}}{\xi_0},\quad\boldsymbol{y}
 =\frac{r_L}{r_S}\frac{\boldsymbol{\eta}}{\xi_0},\quad
  \boldsymbol{d}=\left(1-\frac{r_L}{r_S}\right)\frac{\boldsymbol{\Delta}}{\xi_0},\quad
  w=\frac{r_S\xi_0^2}{r_{LS}r_L}\omega,\quad\psi=\frac{r_Lr_{LS}}{r_S\xi_0^2}\hat\psi
\end{equation}
where we choose $\xi_0$ as
\begin{equation}
  \label{eq:einstein}
  \xi_0=r_L\theta_E,\quad \theta_E=\sqrt{\frac{4M r_{LS}}{r_Lr_S}}.
\end{equation}
$M$ is the mass of the gravitational source, $\xi_0$ and $\theta_E$
represent the Einstein radius and the Einstein angle,
respectively. Using these dimensionless variables,
\begin{eqnarray}
  &F(\boldsymbol{y},\boldsymbol{d})
  =\frac{w}{2\pi i}\int
  d^2x\exp\left[iw\left(\frac{1}{2}(\boldsymbol{x}-\boldsymbol{y}-\boldsymbol{d})^2
      -\psi(\boldsymbol{x})\right)\right],
  \label{eq:ampF}\\
  &\Phi_0(\boldsymbol{y},\boldsymbol{d})=\frac{a_0}{r_S}
\exp\left[iw\left(\frac{r_{LS}}{2r_L}\left(\boldsymbol{y}
-\frac{r_L}{r_{LS}}\boldsymbol{d}\right)^2+\frac{r_{LS}r_{L}}{\xi_0^2}\right)\right].
\end{eqnarray}

\begin{figure}[H]
  \centering
  \includegraphics[width=0.24\linewidth,clip]{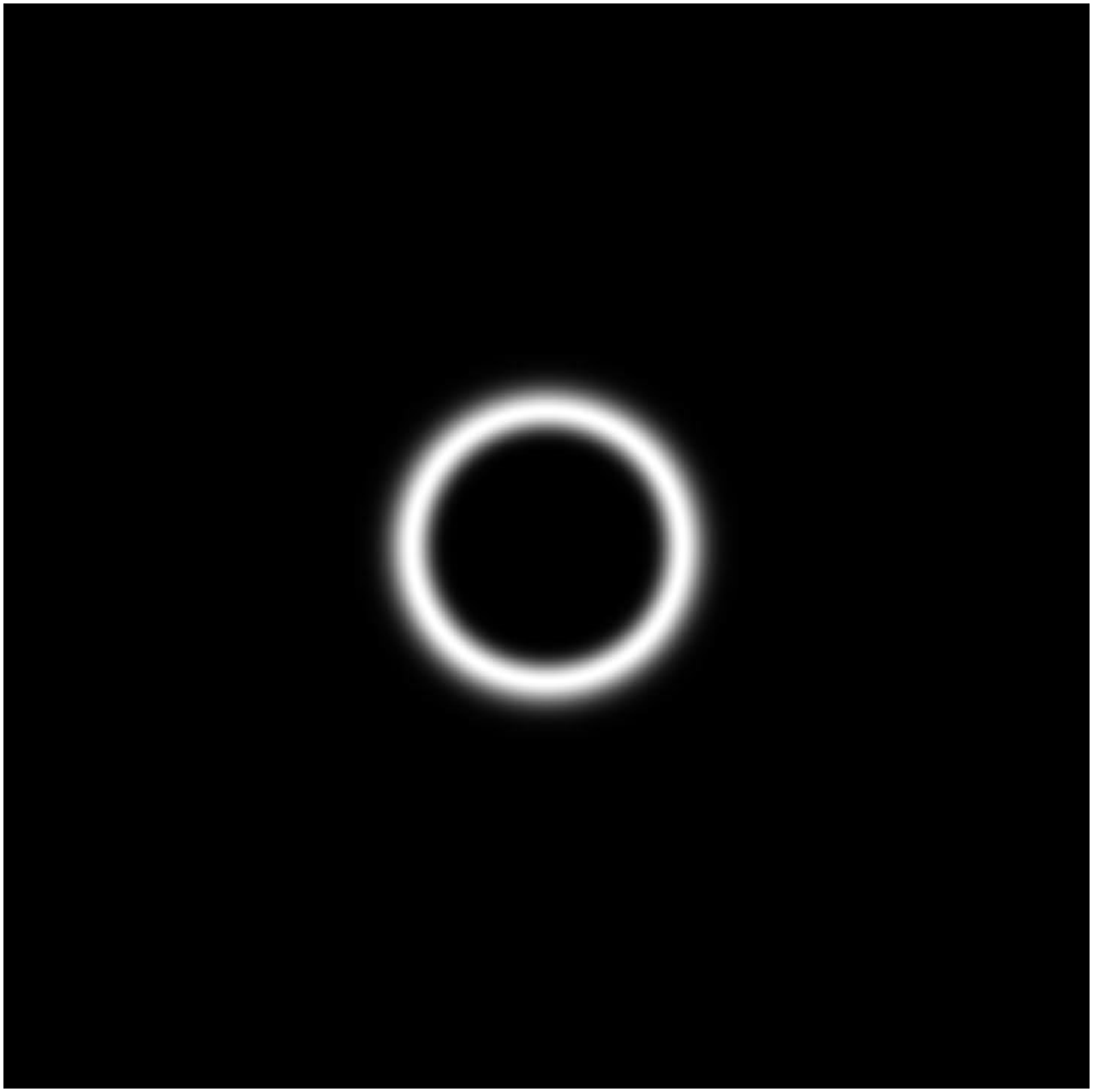}
  \includegraphics[width=0.24\linewidth,clip]{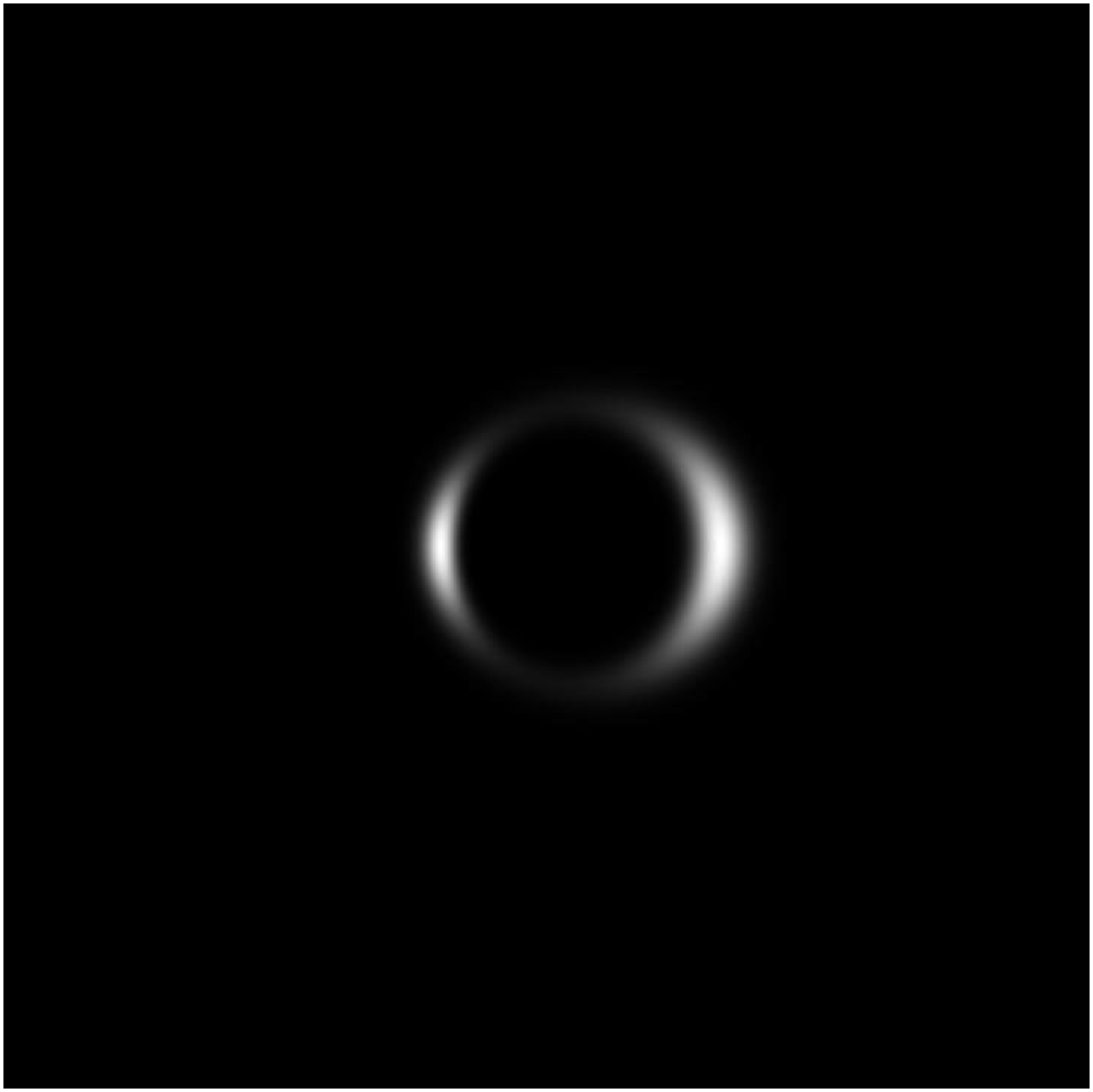}
  \includegraphics[width=0.24\linewidth,clip]{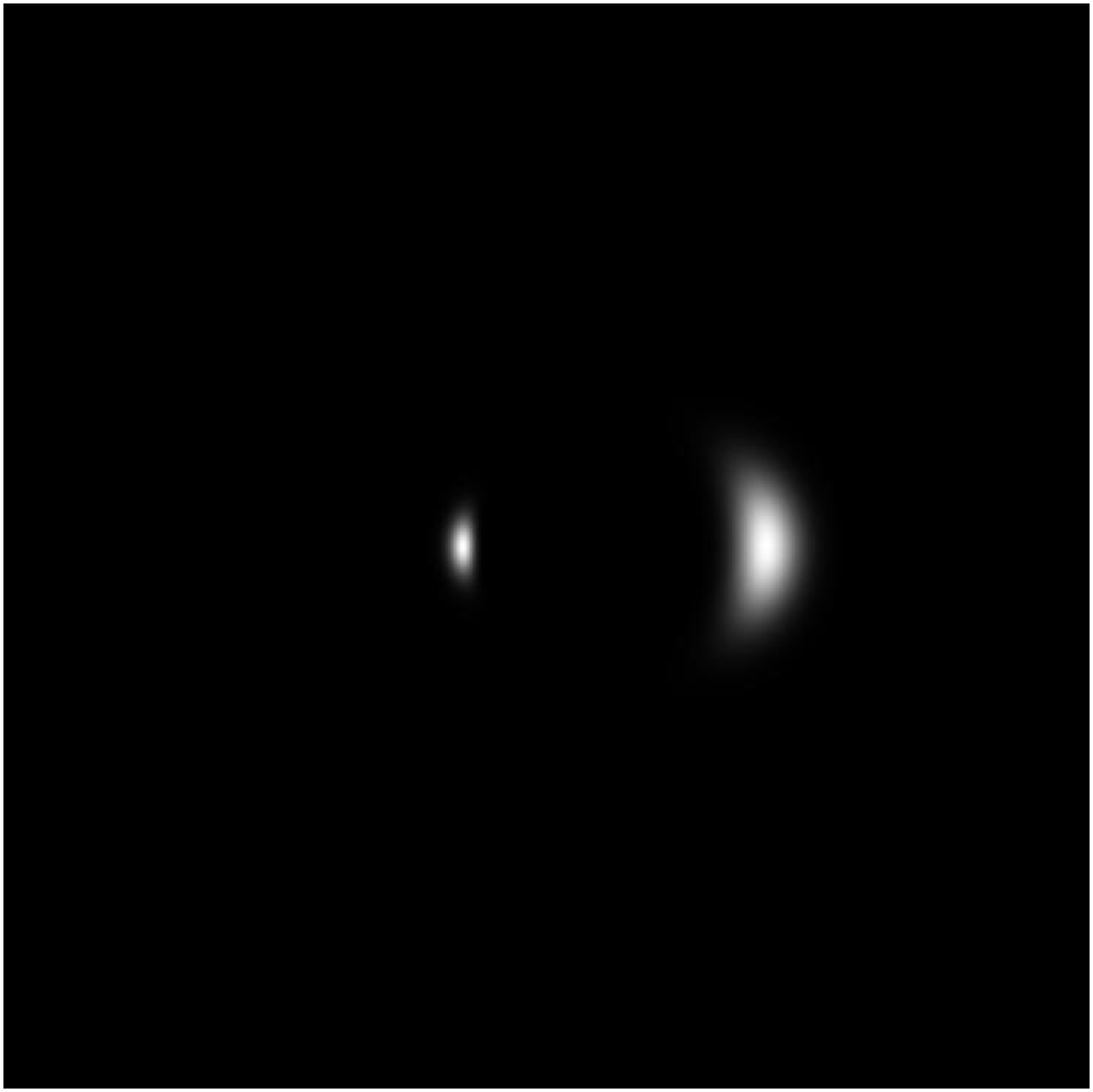}
  \includegraphics[width=0.24\linewidth,clip]{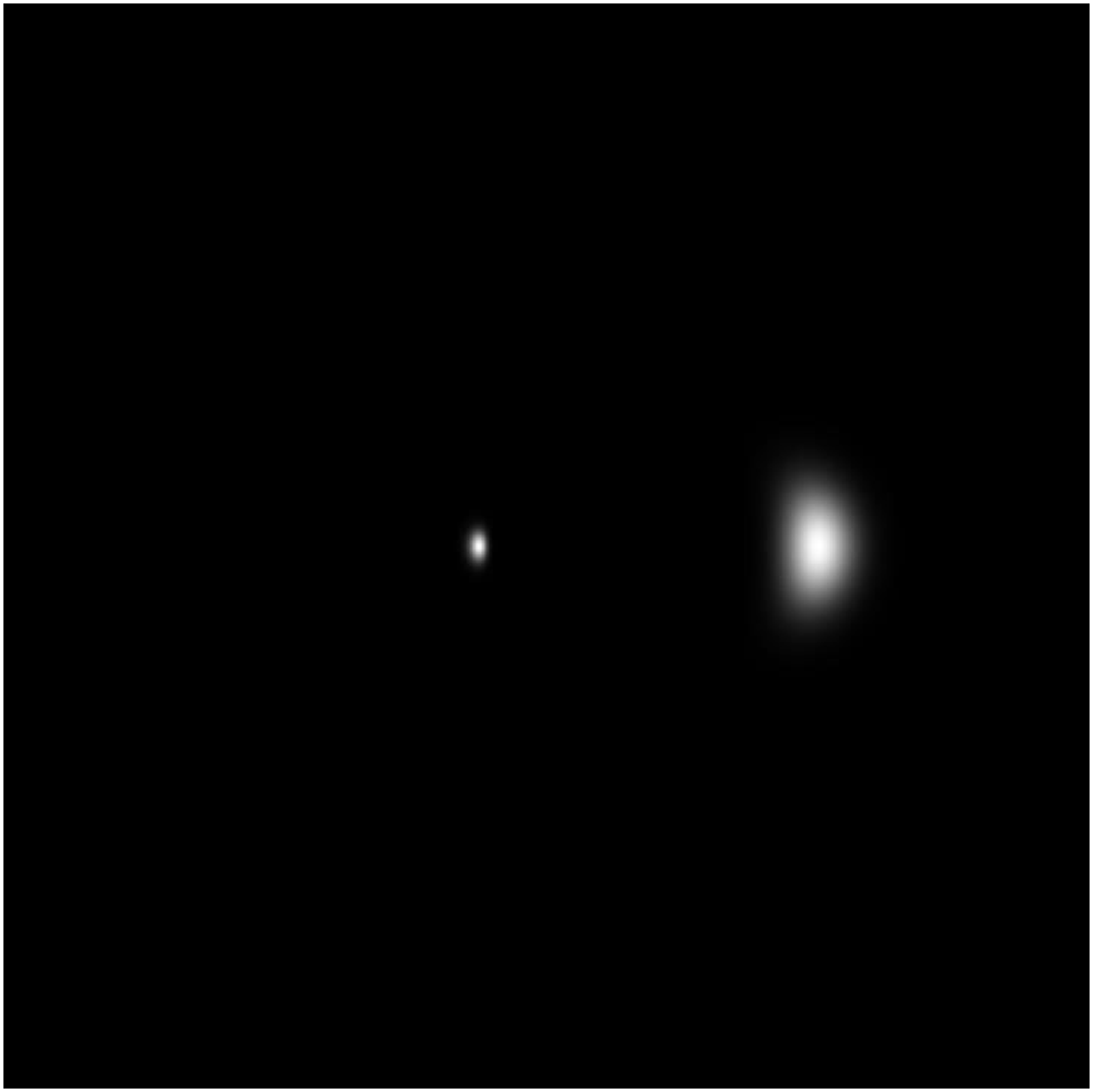}
  \caption{Images of gravitational lensing by a point mass. The source
  is assumed to have the intensity with Gaussian distribution. From
  the left to the right panels, the source positions are $y=0.0, 0.5,
  1.0, 1.5$.}
  \label{fig:image1}
\end{figure}

In the geometrical optics limit $w\gg 1$, the diffraction integral
(\ref{eq:ampF}) can be evaluated around the stationary points of the
phase function in the integrand. The stationary points are determined
by the solution of the following equation:
\begin{equation}
  \boldsymbol{x}-\boldsymbol{y}-\boldsymbol{d}
  -\nabla_{\boldsymbol{x}}\psi(\boldsymbol{x})=0. \label{eq:lens}
\end{equation}
This is the lens equation for gravitational lensing and determines the
location of the image $\boldsymbol{x}$ for given source position $\boldsymbol{y}$. As
the specific model of gravitational lensing, we consider a point mass
as a gravitational source. In this case, the two dimensional
gravitational potential is
\begin{equation}
  \label{eq:pot}
  \psi(\boldsymbol{x})=\ln|\boldsymbol{x}|
\end{equation}
and the deflection angle is given by
\footnote{Using the original variables, the lens equation is $
  \frac{\boldsymbol{\xi}-\boldsymbol{\Delta}}{r_L}-\frac{\boldsymbol{\eta}
    -\boldsymbol{\Delta}}{r_S}
  -\frac{r_{LS}}{r_S}\nabla_{\boldsymbol{\xi}}\hat\psi=0.$}
\begin{equation}
  \al=|\nabla_{\boldsymbol{\xi}}\hat\psi|=\frac{4M}{\xi}.
\end{equation}
For $\boldsymbol{y}=\boldsymbol{d}=0$, the solution of the lens equation
(\ref{eq:lens}) is
\begin{equation}
  |\boldsymbol{x}|=1
\end{equation}
and represents the Einstein ring with the apparent angular radius
$\theta_E$ defined by (\ref{eq:einstein}).  We show an example of
images obtained as solutions of the lens equation (\ref{eq:lens})
(Figure~\ref{fig:image1}).  To produce these images, we have assumed an
extended source with Gaussian distribution of intensity.
\noindent

The wave  property is obtained by evaluating the diffraction
integral (\ref{eq:ampF}). For a point mass lens potential (\ref{eq:pot}), the
integral can be obtained exactly 
\begin{equation}
  \label{eq:F}
  F(\boldsymbol{y})
  =e^{(i/2)w(|\boldsymbol{y}|^2+\ln(w/2))}e^{\pi/4}\Gamma\left(1-\frac{i}{2}w\right){}_1F_1
  \left(1-\frac{i}{2}w, 1, -\frac{i}{2}w|\boldsymbol{y}|^2\right).
\end{equation}
On the observer plane, an interference pattern appears (Figure~3).
\begin{figure}[H]
  \centering
  \includegraphics[width=0.5\linewidth,clip]{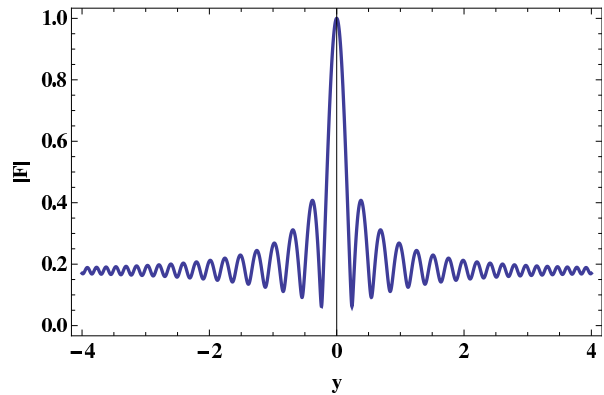}
  \caption{Amplification factor for $w=10$.}
\end{figure}
\noindent
Near $y=0$, the distance between adjacent fringes of the interference
pattern is 
\begin{equation}
  \Delta y\sim\sqrt{\frac{2\pi}{4M\omega}}.
\end{equation}
This fringe pattern is interpreted as interference between double
images of a point source by the gravitational lensing.  The question
we aim to raise in this paper is how the interference pattern on the
observer plane is related to the images of gravitational lensing in
the geometrical optics limit. The wave amplitude on the observer plane
does not make the image of the source and we have to transform the
wave function to extract images. To answer this question, we introduce
a ``telescope'' in the gravitational lensing system and simulate
observation of a star (a point source) using the telescope. With this
setup, it is possible to understand how images of a source are formed
in the framework of wave optics.

\section{Image formation in wave optics}
To establish relation between the interference pattern of the wave and
the images of the source in the gravitational lensing system, we first
consider an image formation system composed of a single convex lens
and review how images of source objects appear in the framework of
wave optics~\cite{SharmaKK:AP:2006}.
\subsection{Image formation by a convex lens}
\begin{figure}[H]
  \centering
  \includegraphics[width=0.4\linewidth,clip]{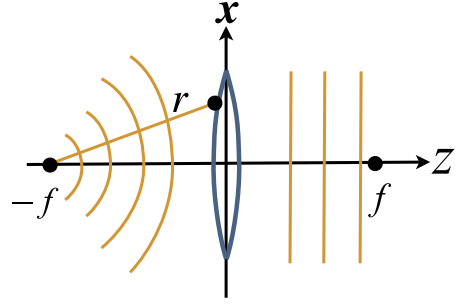}
  \caption{Wave front modification by a convex lens.}
  \label{fig:convex1}
\end{figure}

\begin{figure}[H]
  \centering
  \includegraphics[width=0.5\linewidth,clip]{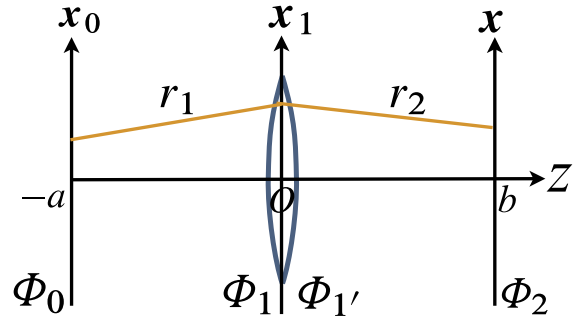}
  \caption{One lens image formation system.}
  \label{fig:lens-config}
\end{figure}

Let us $\Phi_{\mathrm{in}}(\boldsymbol{x})$ is the incident wave from a point
source in front of a thin convex lens and
$\Phi_{\mathrm{t}}(\boldsymbol{x})$ is the transmitted wave by the lens
 (Figure~\ref{fig:convex1}). They are connected by the following relation
\begin{equation}
 \label{eq:conv-lens}
 \Phi_{\mathrm{t}}(\boldsymbol{x})=T(\boldsymbol{x})\Phi_{\mathrm{in}}(\boldsymbol{x}),
\qquad T(\boldsymbol{x})=e^{-i\omega\frac{|\boldsymbol{x}|^2}{2f}}
\end{equation}
where $T(\boldsymbol{x})$ is called a lens transformation function. The action of
a convex lens is to modify the phase of the incident wave. For a point source
placed at $z=-f$ (front focal point), the incident wave and the
transmitted wave are
$$
 \Phi_\mathrm{in}(\boldsymbol{x})=\frac{e^{i\omega r}}{r}\approx
 \frac{e^{i\omega(f+\frac{|\boldsymbol{x}|^2}{2f})}}{f}, \quad
 \Phi_\mathrm{t}(\boldsymbol{x})=T\,\Phi_{\mathrm{in}}=e^{i\omega f},
$$
where we have used $r=\sqrt{f^2+|\boldsymbol{x}|^2}\approx f+|\boldsymbol{x}|^2/(2f)$
assuming $|\boldsymbol{x}|\ll f$. 
Thus, a convex lens converts a spherical wave front to a plane wave front.
Using this action of a convex lens for the incident wave and the
transmitted wave, we can demonstrate the image formation by a convex
lens in the framework of wave optics. Let us consider the
configuration of the lens system shown in
Figure~\ref{fig:lens-config}.
\noindent
We assume the distribution of the source field on the object plane
$z=-a$ as $\Phi_0(\boldsymbol{x}_0)$. Using the Fresnel-Kirchhoff diffraction
formula, the amplitude of the
wave in front of the lens is given by
$$
 \Phi_1(\boldsymbol{x})\propto\int d^2x_0\,\Phi_0(\boldsymbol{x}_0)e^{i\omega r_1}\approx
 \int d^2x_0\,\Phi_0(\boldsymbol{x}_0)
\,e^{i\omega\left(a+\frac{|\boldsymbol{x}_1-\boldsymbol{x}_0|^2}{2a}\right)}
$$
where $r_1$ is the path length from a point on the object plane to a
point on the lens plane and we have assumed $|\boldsymbol{x}_1-\boldsymbol{x}_0|\ll
a$. The amplitude of the wave just behind the lens is given by the
relation (\ref{eq:conv-lens})
$$
 \Phi_{1'}(\boldsymbol{x}_1)=t_L(\boldsymbol{x}_1)
 \,e^{-i\omega\frac{|\boldsymbol{x}_1|^2}{2f}}\Phi_1(\boldsymbol{x}_1)
$$
where $t_L$ is the aperture function of the lens defined by
$t_L(\boldsymbol{x})=1$ for $0\le|\boldsymbol{x}|\le D$ and
$t_L(\boldsymbol{x})=0$ for $D<|\boldsymbol{x}|$. $D$ represents a
radius of the lens.  With the assumption
$|\boldsymbol{x}_1-\boldsymbol{x}|\ll b$, the amplitude of the wave on
the $z=b$ plane behind the lens is
\begin{eqnarray}
\fl \Phi_2(\boldsymbol{x})&\propto\int
d^2x_1\,\Phi_{1'}(\boldsymbol{x}_1)e^{i\omega r_2}
 \nonumber\\
\fl  &\propto\int
  d^2x_0\,d^2x_1\,\Phi_0(\boldsymbol{x}_0)t_L(\boldsymbol{x}_1)
  \,e^{i\frac{\omega}{2a}|\boldsymbol{x}_1-\boldsymbol{x}_0|^2}
  e^{-i\frac{\omega}{2f}|\boldsymbol{x}_1|^2}
  e^{i\frac{\omega}{2b}|\boldsymbol{x}_1-\boldsymbol{x}|^2}
  \nonumber\\
  \fl  &=\int d^2x_0\,d^2x_1\,\Phi_0(\boldsymbol{x}_0)t_L(\boldsymbol{x}_1)
  \exp\left[i\omega\left\{\frac{1}{2}\left(\frac{1}{a}+\frac{1}{b}-\frac{1}{f}\right)
    |\boldsymbol{x}_1|^2-\left(\frac{\boldsymbol{x}_0}{a}
      +\frac{\boldsymbol{x}}{b}\right)\cdot\boldsymbol{x}_1
  \right\}\right] \nonumber\\
\fl  &\qquad\qquad\qquad\qquad\qquad\qquad\qquad\qquad\times
  \exp\left[i\frac{\omega}{2}
    \left(\frac{|\boldsymbol{x}_0|^2}{a}+\frac{|\boldsymbol{x}|^2}{b}
\right)\right].
  \label{eq:phi2}
\end{eqnarray}
 For a value of $b$ satisfying the following relation (the lens equation for a
 convex thin lens), 
\begin{equation}
  \label{eq:lens-eq1}
 \frac{1}{a}+\frac{1}{b}=\frac{1}{f},
\end{equation}
the wave amplitude becomes
\begin{eqnarray}
\fl  \Phi_2(\boldsymbol{x})&\propto
  \int d^2x_0d^2x_1\Phi_0(\boldsymbol{x}_0)t_L(\boldsymbol{x}_1)
  \exp\left[-i\omega\left(\frac{\boldsymbol{x}_0}{a}+\frac{\boldsymbol{x}}{b}\right)
    \cdot\boldsymbol{x}_1\right]\exp\left[i\omega\frac{|\boldsymbol{x}_0|^2}{2a}\right]
  \nonumber \\
  \fl &\propto\int d^2x_0\Phi_0(\boldsymbol{x}_0)
  \left(\frac{2J_1(\omega D|\boldsymbol{x}/b+\boldsymbol{x}_0/a|)}
    {\omega D|\boldsymbol{x}/b+\boldsymbol{x}_0/a|}\right)
\exp\left[i\omega\frac{|\boldsymbol{x}_0|^2}{2a}\right]. \label{eq:phi2Bess}
\end{eqnarray}
For $\omega D\rightarrow\infty$
limit, the Bessel function in (\ref{eq:phi2Bess}) becomes the delta
function and we obtains the following wave amplitude on $z=b$:
\begin{equation}
 \Phi_2(\boldsymbol{x})\propto\int d^2x_0\,\Phi_0(\boldsymbol{x}_0)\times
 \del^2\left[\frac{\boldsymbol{x}_0}{a}+\frac{\boldsymbol{x}}{b}\right]
 =\Phi_0\left(-\frac{a}{b}\boldsymbol{x}\right).
\end{equation}
Thus, a magnified image of the source field appears on the $z=b$
plane. This reproduces the result of image formation in geometric
optics; we have shown that an inverted images with magnification $b/a$
of a source object appears on $z=b$ satisfying the lens
equation~(\ref{eq:lens-eq1}).

If we do not take $\omega D\rightarrow\infty$ limit, due to the
diffraction effect, an image of a point source has finite size on the
image plane called the Airy disk~\cite{SharmaKK:AP:2006}. Its size is
given by
\begin{equation}
  \label{eq:Airy}
  \Delta x_{\mathrm{Airy}}\sim\frac{b\lambda}{D},\quad \lambda=\frac{2\pi}{\omega}.
\end{equation}
This value determines the resolving power of image formation
system. For two point sources at $\boldsymbol{x}_0=-\boldsymbol{d}/2,
\boldsymbol{d}/2$, their separation on the image plane is $bd/a$. To
resolve them, their separation must be larger than the
size of the Airy disk:
\begin{equation}
  \label{eq:resolution}
     \frac{d}{a}>\frac{\lambda}{D}\equiv\theta_0.
\end{equation}
The lefthand side of this inequality is the angular separation of the
sources and $\theta_0$ determines the resolving power of the image
formation system.

\subsection{Image formation in  gravitational lens system}
\begin{figure}[H]
  \centering
  \includegraphics[width=0.8\linewidth,clip]{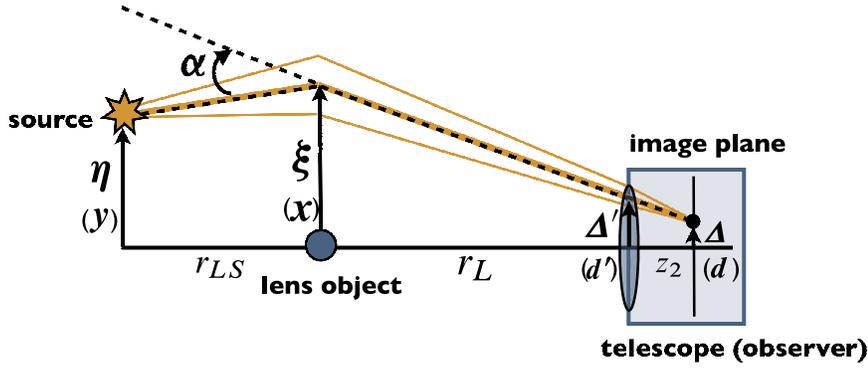}
  \caption{Configuration of a gravitational lens with a convex
    lens system. Thin orange lines represent paths that contribute to
    diffraction integrals.}
  \label{fig:setting2}
\end{figure}

As we have observed that a convex lens can be a device for image
formation in wave optics, we combine it with a gravitational lensing
system and obtain images by gravitational lensing.  We consider a
configuration of the gravitational lens system shown in
Figure~\ref{fig:setting2} and examine how the images of the source
object appear using wave optics.
\noindent
As the source object, we assume a point source of wave. The amplitude of the
wave just in front of a convex lens is
\begin{equation}
 \Phi_{L}(\boldsymbol{y}, \boldsymbol{d}')
 =\Phi_0(\boldsymbol{y},\boldsymbol{d}')F(\boldsymbol{y},\boldsymbol{d}').
\end{equation}
This equation is the same as (\ref{eq:phi}).  After passing through the
convex lens, the wave amplitude on the image plane $z_2$ is given by
\begin{equation}
\fl  \Phi_I(\boldsymbol{\eta},\boldsymbol{\Delta})=\int_{|\boldsymbol{\Delta}'|\leq\Delta_0}
  d^2\Delta'\,\Phi_{L}(\boldsymbol{\eta},\boldsymbol{\Delta}')\exp
  \left[-\frac{i\omega}{2f}\boldsymbol{\Delta}'{}^2\right]\exp\left[\frac{i\omega}{2z_2}
    \left(\boldsymbol{\Delta}-\boldsymbol{\Delta}'\right)^2\right]
\end{equation}
where $\Delta_0$ denotes the aperture of the convex lens.  Using
dimensionless variables, the wave amplitude on the image plane is
\begin{eqnarray}
\fl  \Phi_I(\boldsymbol{y},\boldsymbol{d})=\frac{a_0}{ r_S}\int_{|\boldsymbol{d}'|\leq d_0} 
d^2d' F(\boldsymbol{y}+\boldsymbol{d}')\\
\fl  \quad\times\exp\left[iw\left\{\frac{r_Lr_S}{2r_{LS}}\left(\frac{1}{r_S}
        +\frac{1}{z_2}
        -\frac{1}{f}\right)\boldsymbol{d}'^2-\left(\boldsymbol{y}
+\frac{r_Lr_S}{r_{LS}z_2}\boldsymbol{d}\right)\cdot\boldsymbol{d}'
      +\frac{1}{2}
\left(\frac{r_{LS}}{r_L}\,\boldsymbol{y}^2
  +\frac{r_Lr_S}{r_{LS}z_2}\,\boldsymbol{d}^2\right)\right\}\right]. \nonumber
\end{eqnarray}
If we choose the location of the image plane $z_2$ to satisfy the
following ``lens equation'' for a convex lens,
$$
 \frac{1}{r_S}+\frac{1}{z_2}=\frac{1}{f},
$$
then the wave amplitude on the image plane becomes
\begin{equation}
  \label{eq:waveImage}
\fl
\Phi_I(\boldsymbol{y},\boldsymbol{d})=\frac{a_0}{r_S}\int_{|\boldsymbol{d}'|\leq d_0}
 d^2d'F(\boldsymbol{y}+\boldsymbol{d}')
 \exp\left[-iw\left(\boldsymbol{y}
+\frac{r_Lr_S}{r_{LS}f}\,\boldsymbol{d}\right)\cdot\boldsymbol{d}'\right]
\times
 e^{i(w/2)g(\boldsymbol{y},\boldsymbol{d})}
\end{equation}
where $g=(r_{LS}/r_L)\boldsymbol{y}^2+(r_Lr_S/r_{LS}z_2)\boldsymbol{d}^2$. Thus the
wave amplitude on the image plane is the Fourier transform of the
amplification factor $F$ which gives the interference fringe pattern.
Under the geometrical optics limit $w\gg 1$, $\boldsymbol{x}$ integral in
the amplification factor (\ref{eq:ampF}) can be approximated by the
WKB form
$$
F(\boldsymbol{y}+\boldsymbol{d}')\approx A\,
e^{iw\left[\frac{1}{2}(\boldsymbol{x}_*-\boldsymbol{y}-\boldsymbol{d}')^2
-\psi(\boldsymbol{x}_*)\right]},
$$
where $\boldsymbol{x}_*(\boldsymbol{y},\boldsymbol{d}')$ is the
solution of the lens equation
\begin{equation}
 0=\boldsymbol{x}-\boldsymbol{y}-\boldsymbol{d}'
-\boldsymbol{\nabla}_{\boldsymbol{x}}\psi(\boldsymbol{x})
 \approx\boldsymbol{x}-\boldsymbol{y}-\boldsymbol{\nabla}_{\boldsymbol{x}}
\psi(\boldsymbol{x}).
 \label{eq:lens-eq22}
\end{equation}
We have assumed that the aperture of the convex lens is
sufficiently smaller than the size of the gravitational lensing
system and $|\boldsymbol{d}'|\leq d_0\ll 1$ holds. Then, the wave
amplitude on the image plane is
\begin{eqnarray}
  \fl
  \Phi_I(\boldsymbol{y},\boldsymbol{d})&\propto
  A\,e^{iw\left[\frac{1}{2}(\boldsymbol{x}_*-\boldsymbol{y})^2
-\psi(\boldsymbol{x}_*)\right]}e^{iwg(\boldsymbol{y},\boldsymbol{d})/2}
  \int_{|\boldsymbol{d}'|\leq d_0} d^2d'
  \exp\left[-iw\left(\boldsymbol{x}_*+\frac{r_Lr_S}{r_{LS}f}\,\boldsymbol{d}\right)
    \cdot\boldsymbol{d}'\right] \nonumber \\
\fl  &=A\,e^{iw\left[\frac{1}{2}(\boldsymbol{x}_*-\boldsymbol{y})^2-\psi(\boldsymbol{x}_*)\right]}
  e^{iwg(\boldsymbol{y},\boldsymbol{d})/2}\times
  2\pi d_0^2\,
\frac{J_1(w|\boldsymbol{x}_*+\beta\boldsymbol{d}|d_0)}{w|\boldsymbol{x}_*+\beta
  \boldsymbol{d}|d_0},\quad\beta\equiv\frac{r_Lr_S}{r_{LS}f}.
\end{eqnarray}
For $wd_0\rightarrow\infty$ limit (large lens aperture limit or high
frequency limit), we obtains
\begin{equation}
\label{eq:gl1}
\Phi_I(\boldsymbol{y},\boldsymbol{d})
\propto\del^2\left[\boldsymbol{x}_*(\boldsymbol{y})+\frac{r_Lr_S}{r_{LS}f}
  \,\boldsymbol{d}\right]
\end{equation}
and the image of the point source appears at the following location on
the image plane determined by the lens equation (\ref{eq:lens-eq22}):
\begin{equation}
  \label{eq:gl2}
 \boldsymbol{d}=-\frac{r_{LS}f}{r_Lr_S}\times \boldsymbol{x}_*(\boldsymbol{y}).
\end{equation}
(\ref{eq:gl1}) and (\ref{eq:gl2}) reproduce the same result of image
formation in the geometrical optics (ray tracing) in terms of the wave
optics. This is what we aim to clarify in this paper.
If the lens equation (\ref{eq:lens-eq22}) has multiple solutions
$\boldsymbol{x}_*^{(j)}, j=1,2,\cdots$, the wave amplitude on the image plane becomes
\begin{equation}
\Phi_I(\boldsymbol{y},\boldsymbol{d})\propto \sum_j
A_j\frac{2J_1(w|\boldsymbol{x}_*^{(j)}
+\beta\boldsymbol{d}|d_0)}{w|\boldsymbol{x}_*^{(j)}+\beta
  \boldsymbol{d}|d_0}
\end{equation}
where $A_j$ are constants.

As an example of image formation in a gravitational lensing system
using wave optics, we present the wave optical images of a point
source by the gravitational lensing of a point mass (Figure~7).  They
are obtained by Fourier transformation of the amplification factor $F$
(equation (\ref{eq:waveImage})) and teh lens equation for
gravitational lensing (\ref{eq:lens}) has not been used. This
procedure corresponds to image formation by a convex lens. These
images correspond to images obtained by geometric optics
(Figure~2). We can observe wave effect in these images.
\begin{figure}[H]
  \centering
  \includegraphics[width=0.24\linewidth,clip]{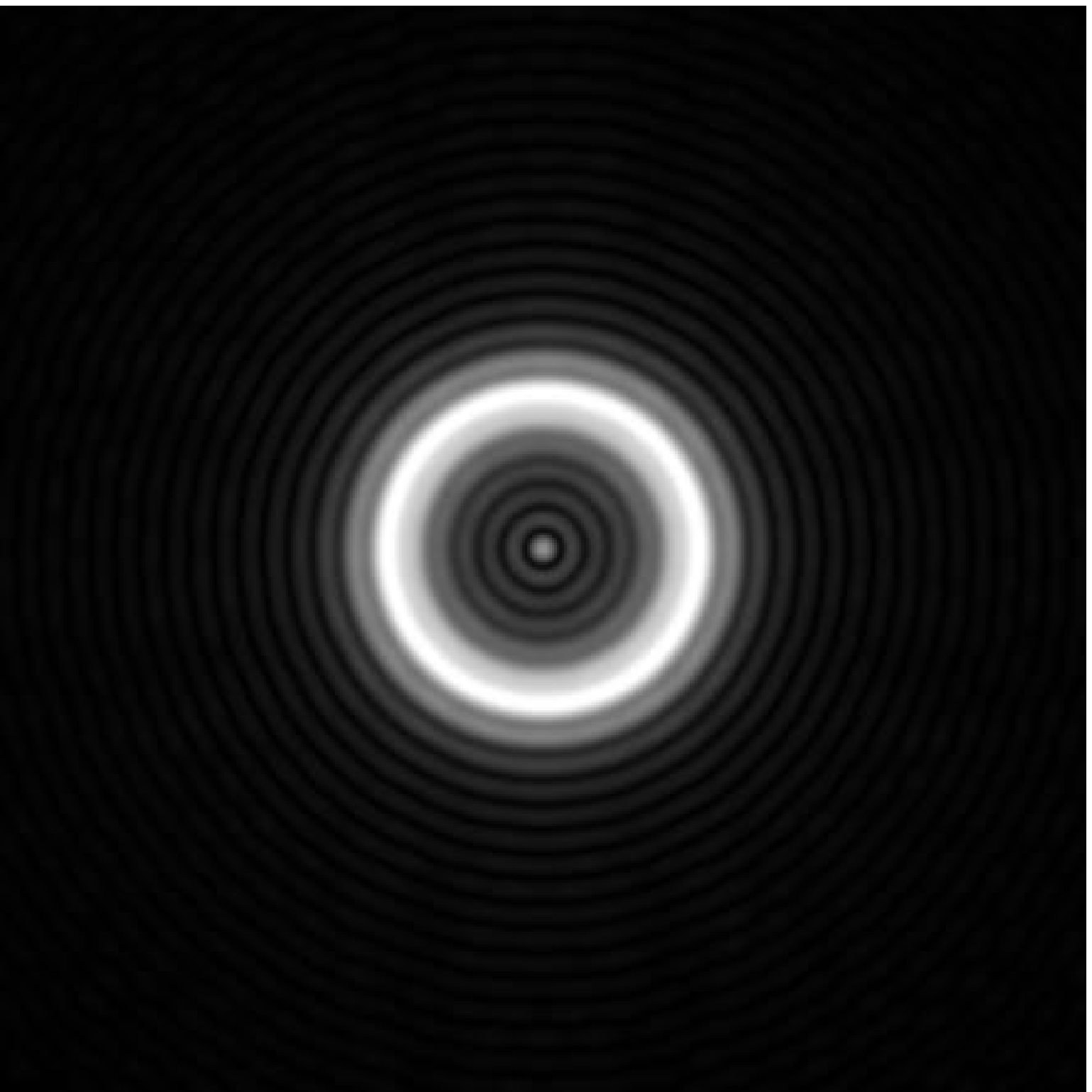}
  \includegraphics[width=0.24\linewidth,clip]{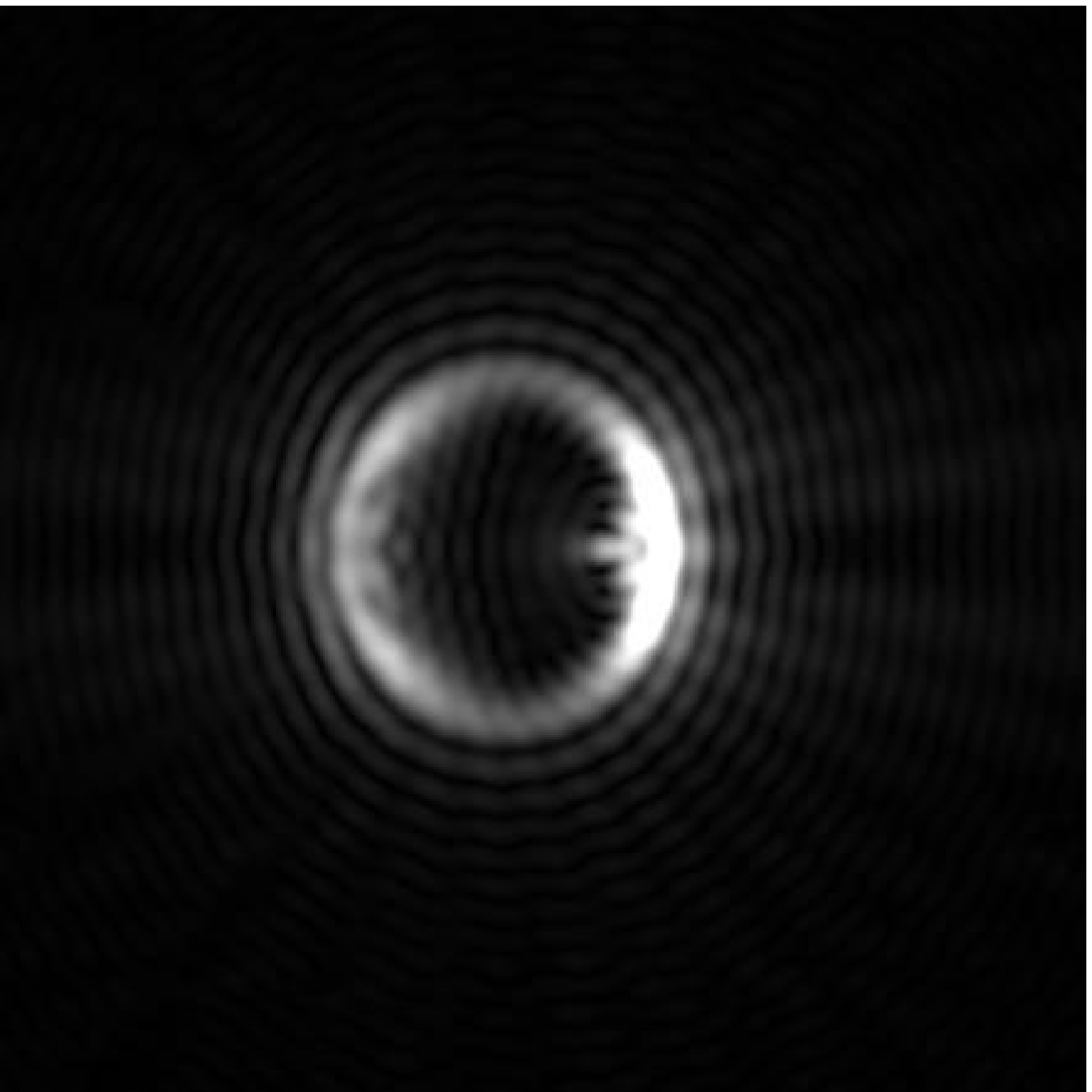}
  \includegraphics[width=0.24\linewidth,clip]{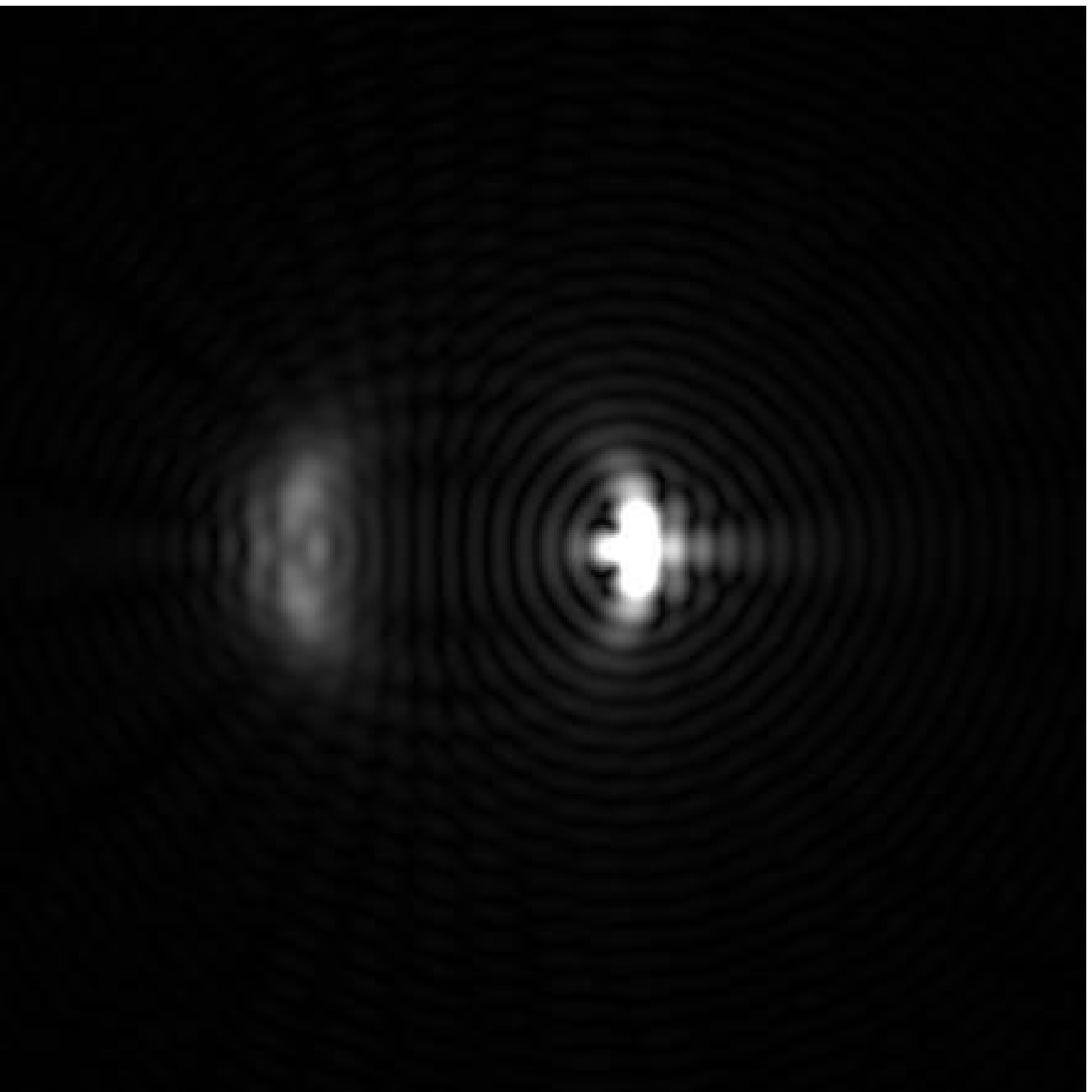}
  \includegraphics[width=0.24\linewidth,clip]{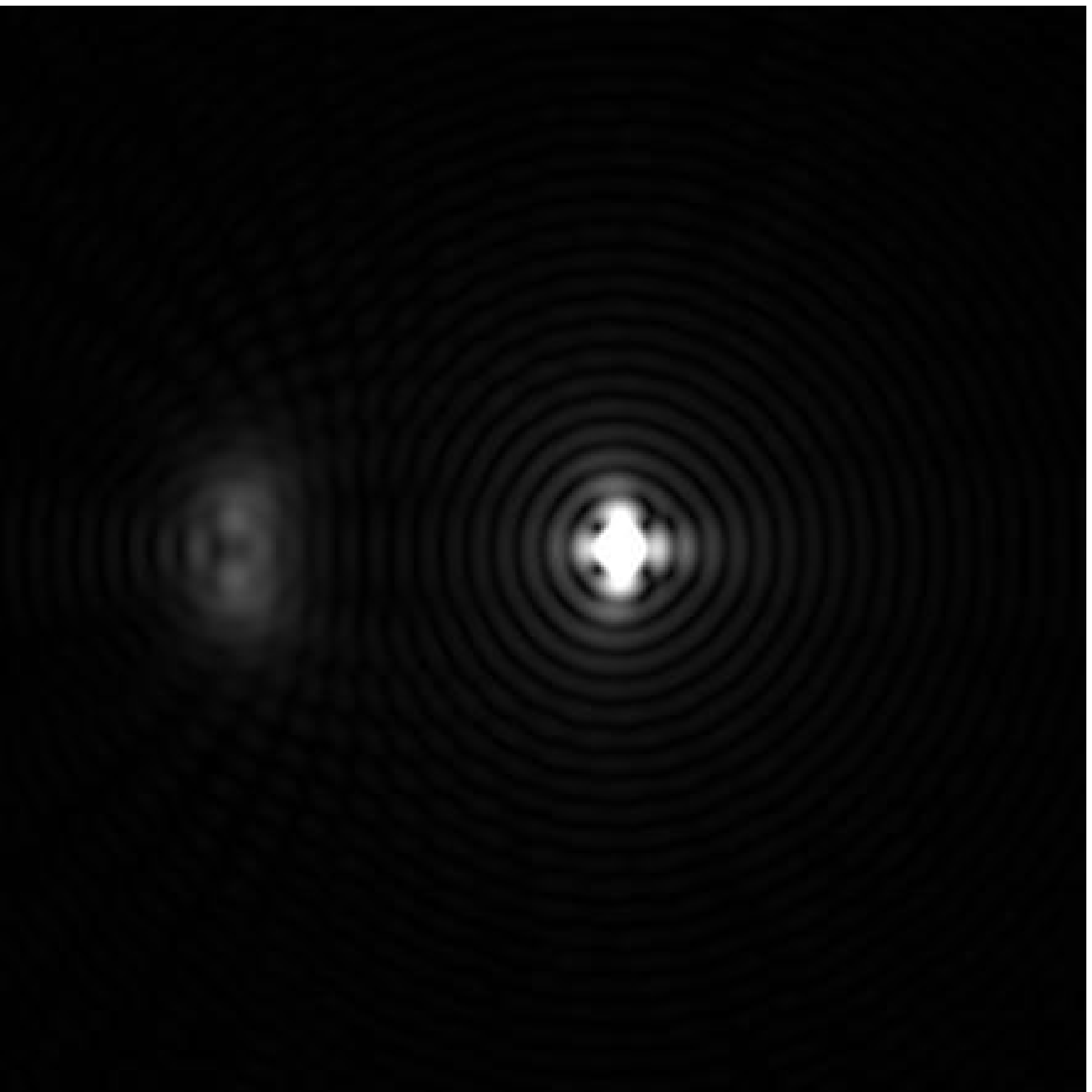}
  \caption{Wave optical images of a point source by the gravitational
    lensing of a point mass. Parameters are $w=40, d_0=0.5$(aperture of
    a convex lens), $y=0,0.5,1,1.5$.}
\end{figure}
\noindent
In each images, we can observe concentric interference pattern which
is caused by finite size of the lens aperture and this is not
intrinsic feature of the gravitational lensing system. We can also
observe radial non-concentric patterns. They are caused by
interference between double images and represent the intrinsic feature
of the gravitational lensing system. For $y=0$ case which corresponds
to the Einstein ring in the geometrical optics limit, we can observe a
bright spot at the center of the ring, which is the result of
constructive interference and does not appear in geometric optics.
For sufficiently large values of $wd_0$, the wave amplitude at the
observer coincides with the result obtained by geometric optics.

\section{Summary}

We investigated image formation in gravitational lensing system based
on wave optics. Instead of using a ray tracing method, we obtained
images directly from wave functions at the observer without using a
lens equation of gravitational lensing.  For this purpose, we
introduced a ``telescope'' with a single convex thin lens, which acts
as a Fourier transformer for the interference pattern formed at an
observer.  The analysis in this paper relates the wave amplitude and
images of the gravitational lensing directly. In the geometric optics
limit of waves, images by lens systems are obtained by a lens equation
which determines paths of each light rays. As light rays are
trajectories of massless test particles (photon), expressing image in
terms of wave is to express particle motion in terms of waves.

As an application and extension of analysis presented in this paper,
we plan to investigate gravitational lensing by a black hole and
obtain wave optical images of black holes. This subject is related to
observation of black hole
shadows~\cite{FalckeH:AJ528:2000,MiyoshiM:PTPS155:2004}. As the
apparent angular sizes of black hole shadows are so small, the
diffraction effect on images are crucial to resolve black hole shadows
in observation using radio interferometer. For SgrA${}^*$, which is
the black hole candidate at Galactic center, the apparent angular size
of its shadow is estimated to be $\sim 30\mu$ arc seconds and this
value is the largest among black hole candidates. For a sub-mm VLBI
with a baseline length $D$, using (\ref{eq:resolution}), the condition
to resolve the shadow becomes
$$
 D>1000~\mathrm{km},
$$
and this requirement shows the possibility to detect the black hole
shadow of SgrA${}^*$ using the present day technology of VLBI
telescope. Thus, analysis of black hole shadows based on wave optics
is an important task to evaluate detectability of shadows and determination of
black hole parameters via imaging of black holes.

The topic of wave optical image formation in black hole spacetimes
belongs to a classical problem of wave scattering in black hole
spacetimes~\cite{FuttermanJAH:CUP:1988}.  As is well known, waves
incident to a rotating black hole are amplified by the
superradiance~\cite{FrolovYP:1998} due to dragging of spacetimes. This
effect enables waves to extract the rotation energy of black holes. On
the other hand, it is known that particles can also extract the
rotation energy of black holes via so called Penrose process.  By
investigating images of scattered waves by a rotating black hole, we
expect to find out new aspect or interpretation of phenomena
associated with superradiance in connection with the Penrose process.

\ack
This work was supported in part by the JSPS Grant-In-Aid for
Scientific Research (C) (23540297). The author thanks all member of
``black hole horizon project meeting'' in which the preliminary version of
this paper was presented.

\end{document}